\documentclass{aa}
\usepackage{graphicx}
\usepackage{txfonts}
\usepackage{natbib}
\begin{document}

   \title{Radial metallicity profiles for a large sample of galaxy clusters
          observed with \emph{XMM-Newton}}

%   \subtitle{I. Overviewing the $\kappa$-mechanism}

   \author{A. Leccardi
          \inst{1,2}
          \and
          S. Molendi\inst{2}
          }

%   \offprints{G. Wuchterl}

   \institute{Universit\`a degli Studi di Milano, Dip. di Fisica, via Celoria 16,
              I-20133 Milano, Italy
%             \email{wuchterl@amok.ast.univie.ac.at}
         \and
             INAF-IASF Milano, via Bassini 15, I-20133 Milano, Italy\\
%             \email{c.ptolemy@hipparch.uheaven.space}
%             \thanks{The university of heaven temporarily does not accept e-mails}
             }

   \date{Received May 5, 2008; accepted June 4, 2008}

  \abstract
   {}
    {We measured radial metallicity profiles for a sample of $\approx 50$ hot,
intermediate redshift galaxy clusters, selected from the \emph{XMM-Newton} archive.}
   {As in our previous paper, we used background modeling rather than background
subtraction, and the Cash statistic rather than the $\chi^2$.
This method requires a careful characterization of all background components.
We also performed montecarlo simulations to assess systematic effects.}
   {The mean metallicity profile shows a peak in the center, and gently declines out
to 0.2~$R_{180}$.
Beyond 0.2~$R_{180}$ the metallicity is $\approx 0.2$ solar and, at variance with
recently published expectations based on simulations, consistent with being flat.
We find no evidence of profile evolution from $z = 0.1$ to $z = 0.3$.
When comparing our mean profile to those obtained by recent works with \emph{BeppoSAX}
and \emph{Chandra}, we find remarkable agreement over the entire radial range.}
   {}


   \maketitle
%
%________________________________________________________________

\section{Introduction} \label{sec: intro}
Clusters of galaxies are the most massive gravitationally bound systems in the universe.
They are permeated by the hot, X-ray emitting intra-cluster medium (ICM), which represents
the dominant baryonic component.
The mean metallicity of the ICM is found to be roughly half of the solar value \citep{lowe04},
therefore a substantial fraction of plasma is not of primordial origin.
As heavy elements are only produced in stars, the processed material must have been ejected
by cluster galaxies into the ICM.
In rich clusters the plasma reaches temperatures of several $10^7$~K and emits X-rays mainly
via thermal bremsstrahlung.
At such temperatures, most elements are either fully ionized or in a high ionization state.
The strong emission lines due to the transitions to the n=1 level of the H-like and He-like
ions of Iron around 6.7~keV (rest frame energy) are the most prominent features in X-ray
spectra of hot clusters.
In fact, for hot clusters as those belonging to our sample, the measurement of the
metallicity is a measurement of the iron abundance.

Spatially resolved analysis of metal abundance in clusters has become possible only recently,
first with \emph{ASCA} and \emph{BeppoSAX} and then with \emph{Chandra} and \emph{XMM-Newton}.
These measurements (e.g. \citealp{grandi01,grandi04,tamura04,baldi07}) have shown that
abundance gradients are common features in clusters of galaxies: most clusters show a peak
of metallicity in the center and a gentle decline outward.

In this paper we present radial metallicity profiles from the same sample of hot intermediate
redshift clusters analyzed in our previous paper \citep{leccardi08}, hereafter Paper~I.
We used the same novel data analysis technique, which is extensively described in Paper~I and
summarized in Sect.~\ref{sec: analysis}.
In Sect.~\ref{sec: mean prof} we show the mean metallicity profile, in
Sect.~\ref{sec: evolution} search for an indication of the evolution of the metallicity
with redshift, and in Sects.~\ref{sec: sim} and \ref{sec: comp} compare our mean profile
with those obtained from hydrodynamic simulations and from previous observational works.
In the Appendix~\ref{app} we describe a systematic effect, which could affect the measurement
of the metallicity especially for the case of low metallicity, statistically poor spectra.

Quoted confidence intervals are 68\% for one interesting parameter (i.e. $\Delta$C~=~1),
unless otherwise stated.
All results are given assuming a $\Lambda$CDM cosmology with $\Omega_\mathrm{m} = 0.3$,
$\Omega_\Lambda = 0.7$, and $H_0$~=~70~km~s$^{-1}$~Mpc$^{-1}$.

\section{The sample} \label{sec: sample}
We selected from the \emph{XMM-Newton}\footnote{An ESA science mission with instruments and
contributions directly funded by ESA Member States and NASA} archive a sample of hot
($\mathrm{k}T > 3.3$~keV), intermediate redshift ($0.1 \lesssim z \lesssim 0.3$), and high galactic
latitude ($|b| > 20^\circ$) clusters of galaxies.
We retrieved from the public archive all observations of clusters satisfying the above
selection criteria, performed before March 2005 (when the CCD6 of EPIC-MOS1 was switched
off\footnote{http://xmm.esac.esa.int/external/xmm\_news/items/MOS1-CCD6/\\
index.shtml}) and available at the end of May 2007.
We excluded from the sample observations that are highly affected by soft proton flares
and observations of clusters that show evidence of recent and strong interactions.
In Table~3 of Paper~I we list the 48 observations that survived our selection criteria and
report cluster physical properties.
Each observation was performed by using THIN1 or MEDIUM filters, and its total
(i.e. MOS1+MOS2) exposure time is greater than 16~ks.

\section{Data analysis}  \label{sec: analysis}
In this section we only recall the major steps of our data analysis procedure and refer to
Paper~I for a detailed description.
We stress that in our procedure we used only EPIC-MOS data, because a robust characterization
of EPIC-pn background was not possible, mainly due to the small regions of the detector
outside the field of view, and to the non-negligible fraction of out-of-time events (see
Appendix B in Paper~I for further details).

\subsection{Spectra preparation}  \label{sec: spec prep}
Observation data files (ODF) were retrieved from the \emph{XMM-Newton} archive and processed
in a standard way with the Science Analysis System (SAS) v6.1.

The soft proton cleaning was performed using a double filtering process, first in a hard
(10-12~keV) and then in a soft (2-5~keV) energy range.
We filtered the event file according to \verb|PATTERN| and \verb|FLAG| criteria, and excluded
the ``bright'' CCDs, i.e. CCD-4 and CCD-5 for MOS1 and CCD-2 and CCD-5 for MOS2
(see Appendix~A in Paper~I for a discussion).
Bright point-like sources were detected, using a procedure based on the SAS task
\verb+edetect_chain+, and excluded from the event file.

The $R_\mathrm{SB}$ indicator, i.e. the ratio between surface-brightness calculated inside
and outside the field of view (see Eq.~1 in Paper~I), allowed us to quantify the amount of
the quiescent soft proton (QSP) component survived the double filtering process.
Values of $R_\mathrm{SB}$ roughly span the range from 1.0 (negligible contamination) to
1.5 (high contamination).
Since the observation 0084230401 of Abell~267 is extremely polluted by QSP
($R_\mathrm{SB}=1.8$), we excluded it from the sample.

The cluster emission is divided in 10 concentric rings (namely 0$^\prime$-0.5$^\prime$,
0.5$^\prime$-1$^\prime$, 1$^\prime$-1.5$^\prime$, 1.5$^\prime$-2$^\prime$,
2$^\prime$-2.75$^\prime$, 2.75$^\prime$-3.5$^\prime$, 3.5$^\prime$-4.5$^\prime$,
4.5$^\prime$-6$^\prime$, 6$^\prime$-8$^\prime$, and 10$^\prime$-12$^\prime$) at fixed
angular radii, to maintain under control systematics related to the detectors (see the
Appendices of Paper~I).
The most external ring was used to estimate background parameters only.
The width of most of the rings is limited by the PSF of the \emph{XMM-Newton} telescopes.
The center of the rings was determined by surface-brightness isocontours at large radii
and is usually, but not necessarily, coincident with the X-ray emission peak.
We prefer that azimuthal symmetry be preserved at large radii, where we are interested in
characterizing profiles, at the expense of central regions.
For each instrument (i.e MOS1 and MOS2) and each ring, we accumulated a spectrum and generated
an effective area (ARF).
For each observation we generated one redistribution function (RMF) for MOS1 and one for MOS2.
We performed a minimal grouping to avoid channels with no counts, as required by the
Cash statistic.

\subsection{Spectral analysis} \label{sec: sp ana}
The spectral analysis followed a substantially new\footnote{A somewhat similar procedure was
already used by \cite{stanford01}} approach: we used the background modeling, rather than
the subtraction, and the Cash statistic rather than the $\chi^2$.
We fit spectra with an absorbed thermal (WABS*MEKAL in XSPEC v11.3\footnote{http://heasarc.nasa.gov/docs/xanadu/xspec/xspec11/index.html})
plus background model in the 0.7-10.0~keV energy band, which represents a good trade off
between statistical quality and level of systematics.
The details of the background model are reported in the Appendices of Paper~I.
To model the background, a careful characterization of all its components is mandatory.
Ideally, one would like to estimate background parameters in the same region and at the same
time as the source.
Since this was not possible, we estimated background parameters in the external
10$^\prime$-12$^\prime$ ring and rescaled them in the inner rings, by making reasonable
assumptions on their spatial distribution tested by analyzing blank-field observations.

We fit spectra in internal rings leaving the temperature and the normalization free to vary;
the metallicity was constrained between $\pm5 \; Z_\odot$ (see the discussion in Appendix A).
The redshift was constrained between $\pm$7\% of the optical measurement in the two innermost
rings and, in the other rings, was fixed to the average value of the first two rings, by
considering independently MOS1 and MOS2 spectra.
The main reason for our choice is to allow for EPIC calibration uncertainties, and for
possible discrepancies between X-ray and optical derived redshift values.
Typical shift values are on the order of 2\%.
Finally, we produced metallicity profiles for each cluster, by computing the average over the
two MOS values.

\section{Metallicity profiles}

\subsection{The mean profile} \label{sec: mean prof}
\begin{figure}
  \centering
  \resizebox{80mm}{!}{\includegraphics[angle=0]{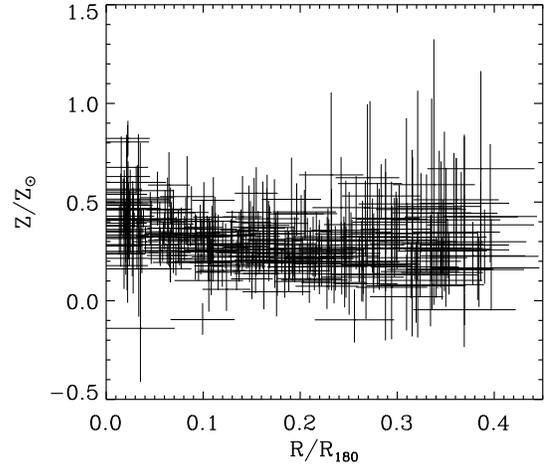}} \\
  \caption{Radial metallicity profiles for all clusters in our sample.
  Abundances are expressed in \cite{anders89} solar values and radii in
  units of $R_{180}$.}
  \label{fig: prof all}
\end{figure}
In Fig.~\ref{fig: prof all} we show radial metallicity profiles for all clusters in our sample.
All our metallicity measurements are relative to the solar values published by \cite{anders89}.
Even if these have been superseded by more recent values \citep{grevesse98,asplund05}, they
allow straightforward comparison with most of the literature \citep{grandi01,balestra07,baldi07}.
\cite{grevesse98} and \cite{asplund05} introduced a 0.676 and 0.60 times lower iron solar
abundance respectively, while other elements are substantially unchanged.
A simple scaling by 0.676 and 0.60 converts measures from the \cite{anders89} iron abundance
to the \cite{grevesse98} and \cite{asplund05} abundances.

The radius is rescaled by $R_{180}$, i.e. the radius encompassing a spherical density contrast
of 180 with respect to the critical density.
We compute $R_{180}$ from the mean temperature, $\mathrm{k}T_\mathrm{M}$, and the redshift, $z$,
as in Paper~I and in \cite{arnaud05}:
\begin{equation}
\label{eq: r180}
R_{180}=1780\,\left(\frac{\mathrm{k}T_\mathrm{M}}{5\,\mathrm{keV}}\right)^{1/2}
h(z)^{-1} \; \mathrm{kpc} ,
\end{equation}
where $h(z)=(\Omega_\mathrm{M}(1+z)^3+\Omega_\Lambda)^{1/2}$.
$R_{180}$ is a good approximation to the virial radius in an Einstein-De~Sitter universe
and has been largely used to rescale cluster radial properties (e.g.
\citealp{grandi04,vikh05,pratt07,leccardi08}).
The profiles are limited to $\approx 0.4 \; R_{180}$ because beyond this radius the
source-to-background count rate ratio is too small and the measurements are unreliable
(see Sect~3.2.2 in Paper~I).

\begin{figure}
  \centering
  \resizebox{80mm}{!}{\includegraphics[angle=0]{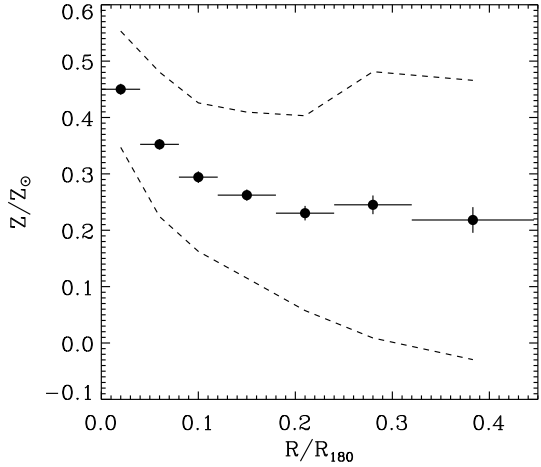}} \\
  \caption{Mean metallicity profile averaged over all clusters. Abundances
  are expressed in \cite{anders89} solar values and radii in units of
  $R_{180}$. The dotted lines show the one-sigma scatter of the values
  around the average.}
  \label{fig: mean prof}
\end{figure}
\begin{table}
  \caption{Mean metallicity profile.}
  \label{tab: values}
  \centering
  \begin{tabular}{ll}
    \hline \hline
    Ring$^a$ & Metallicity$^b$ \\
    \hline
    0.00-0.04 & 0.450$\pm$0.010 \\
    0.04-0.08 & 0.352$\pm$0.009 \\
    0.08-0.12 & 0.294$\pm$0.010 \\
    0.12-0.18 & 0.262$\pm$0.010 \\
    0.18-0.24 & 0.230$\pm$0.013 \\
    0.24-0.32 & 0.245$\pm$0.017 \\
    0.32-0.45 & 0.218$\pm$0.023 \\
    \hline
  \end{tabular}
  \begin{list}{}{}
    \item[Notes:] $^a$ in units of $R_{180}$; $^b$ in solar units \citep{anders89}.
  \end{list}
\end{table}
In Table~\ref{tab: values} and in Fig.~\ref{fig: mean prof} we report the mean
metallicity profile binned in units of $R_{180}$; these values have been computed by
performing, for each new bin, a weighted average of metallicity values in the
original bins which have a non-zero intersection with the new bin.
The weight is the product of two components: one is the inverse squared errors, the other
depends on the intersection between the original bin and the new one.
If the original bin is totally included into the new one, the weight is equal to one.
If the original bin has only a partial intersection with the new one, the
weight is the fraction of the original bin that belongs to the new one.
Possible blurring effects associated either to the original binning in angular units or to
the recasting in units of $R_{180}$ should be minimal.
As far as the original binning is concerned, we note that the size of central bins (i.e.
$30^{\prime\prime}$) is comparable to the \emph{XMM-Newton} PSF.
As far as the recasting is concerned, the one we show in Fig.~\ref{fig: mean prof} is the
result of various trials specifically aimed at reaching the best compromise between
resolution and statistical quality; moreover, the size of central bins
(i.e. $0.04 \; R_{180}$) corresponds to $\approx 30^{\prime\prime}$ for clusters at
$z \approx 0.2$ with a temperature of 7~keV.

\cite{balestra07} and \cite{maughan08} have adopted a procedure that is alternative to ours
to estimate mean metal abundances for a sample of clusters.
These authors performed a simultaneous spectral fit, leaving temperature and normalization
free to vary for each object and using a unique metallicity value for all clusters in each
redshift bin.
We note that when modeling the
background, as we do, a joint fit is infeasible, because of
the large number of model parameters.
Moreover, we
are not aware of any detailed work that investigates the impact of systematic
errors,
possibly affecting individual measurements, on the final result of a joint fit.

The mean metallicity is $0.45 \; Z_\odot$ in the center and decreases out to
$\approx 0.2 \; R_{180}$; beyond this radius the profile is consistent with being flat,
a fit with a constant for $R>0.2 \; R_{180}$ gives $Z=0.23\pm0.01 \; Z_\odot$.
The profiles show a large scatter, which is mostly of statistical origin.
In the central regions ($R<0.2 \; R_{180}$) we find an intrinsic scatter of 22~$\pm$~2\%
related to the presence of cool core clusters; in the outer regions ($R>0.2 \; R_{180}$)
the intrinsic scatter is only 14~$\pm$~8\% (i.e. $\approx 0.03 \; Z_\odot$), the same
order of magnitude as our systematics (see Appendix~\ref{app}).
Past works (e.g. \citealp{grandi01}) have shown that the abundance profiles of cool core
and non cool core clusters (see Sect.~4 in Paper~I) differ in the central regions.
We found qualitatively similar results, but we choose to extensively address this important
issue in a forthcoming paper (Leccardi et al., in prep.), where we also compare our results
with those obtained from a local sample.

\subsection{Redshift evolution} \label{sec: evolution}
We divided clusters in our sample into two groups to investigate a possible profile
evolution with redshift: near (distant) clusters are characterized by a redshift lower
(greater) than 0.2.
\begin{figure}
  \centering
  \resizebox{80mm}{!}{\includegraphics[angle=0]{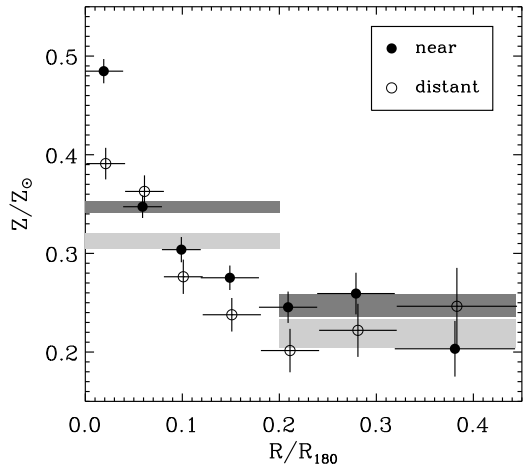}} \\
  \caption{Mean metallicity profiles for near ($z<0.2$, filled circles)
  and distant ($z>0.2$, empty circles) clusters. The dark and the light
  shaded regions indicate the average metallicity within one-sigma
  uncertainties below and beyond $0.20 \; R_{180}$ for near and distant
  clusters, respectively. The radii have been slightly offset in the plot
  for clarity.}
  \label{fig: cfr z}
\end{figure}
In Fig.~\ref{fig: cfr z} we show the mean metallicity profiles for the two groups.
The single points are consistent within one- or two-sigma, except for the core region
where the spatial resolution plays an important role.
In our procedure we fixed the size of the central ring to 30$^{\prime\prime}$, which
corresponds to $\approx 0.03 \; R_{180}$ for nearest ($z \approx 0.1$) and to
$\approx 0.07 \; R_{180}$ for most distant ($z \approx 0.3$) objects; thus, the metallicity
peak for distant clusters is blurred over a larger region.
The three-sigma discrepancy in the region within 0.2~$R_{180}$ (see shaded regions in
Fig.~\ref{fig: cfr z}) is most likely due to a different fraction of cool core and non cool
core clusters within the two subsamples (Leccardi et al., in prep.).

In the outer regions, if the metallicity is allowed to vary between 0 and $5 \; Z_\odot$ as
commonly done, we measure a discrepancy between two profiles of $0.061\pm0.018 \; Z_\odot$
with a more than three-sigma significance (see Appendix~\ref{app}).
Conversely, if the metallicity is allowed to vary between $\pm5 \; Z_\odot$ as in our data
analysis procedure (see Sect.~\ref{sec: sp ana}), near and distant clusters have a mean
metallicity of $0.248\pm0.011 \; Z_\odot$ and $0.219\pm0.015 \; Z_\odot$, respectively.
For this case, the discrepancy (i.e. $\approx 1.5$~sigma, see shaded regions in
Fig.~\ref{fig: cfr z}) is consistent with a purely statistical fluctuation, and is the same
order of magnitude as our systematics.
Summarizing, we find no evidence of metallicity profile evolution from $z = 0.1$ to $z = 0.3$.
We stress that the systematic effect described in the Appendix~\ref{app}, which affects in
particular low metallicity, statistically poor spectra, if unaccounted for, could cause a
false detection of the metallicity evolution.

Two recent works \citep{balestra07,maughan08} have investigated the evolution in the iron
content of the ICM, by analyzing data from the \emph{Chandra} archive.
As mentioned in Sect.~\ref{sec: mean prof}, these authors performed a simultaneous spectral
fit, leaving temperature and normalization free to vary for each object and using a unique
metallicity value for all clusters in each redshift bin.
Unfortunately, we cannot compare our results with those obtained by \cite{balestra07},
because they considered different regions from cluster to cluster.
Instead, \cite{maughan08} analyzed the region within $R_{500}$ ($\approx 0.6 \; R_{180}$)
with and without the core region (i.e. 0.15~$R_{500}$) and obtained respectively
$Z \approx 0.4 \; Z_\odot$ and $Z \approx 0.35 \; Z_\odot$ between $z = 0.1$ and $z = 0.3$.
When analyzing roughly the same regions for our clusters, we obtain $Z = 0.32 \; Z_\odot$
and $Z = 0.26 \; Z_\odot$ respectively, with negligible uncertainties; i.e., a significantly
lower ($\approx 20\%$) mean metallicity.
A possible explanation for this discrepancy may be related to different weights on the
averaging procedure: we averaged weighting over the inverse squared errors, while \cite{maughan08}
measured metal abundances from individual spectra extracted from the entire region of interest.

\subsection{Comparison with hydrodynamic simulations} \label{sec: sim}
We compare our (hereafter LM08) mean metallicity profile with the one (hereafter F08) derived
from hydrodynamic simulations of four relaxed clusters by \cite{fabjan08}.

The simulations are performed using the hydrodynamical TREE-SPH code GADGET-2 \citep{springel05}
with the implementation of chemical enrichment by \cite{tornatore07}.
The authors used the emission-weighted definition of metallicity, with emissivity of each gas
particle computed in the 0.5-10.0~keV energy band.
In principle, for a comparison with observational data, one should extract synthetic spectra
from the simulated clusters and then measure the metallicity by fitting these spectra with a
single-temperature and single-metallicity plasma model (e.g. the MEKAL model in XSPEC).
A recent work presented by \cite{rasia08} showed that, at least for Iron, the emission-weighted
estimator of the metallicity gives results quite close (within about 10\%) to those obtained
from the spectral-fitting analysis.

\begin{figure}
  \centering
  \resizebox{80mm}{!}{\includegraphics[angle=0]{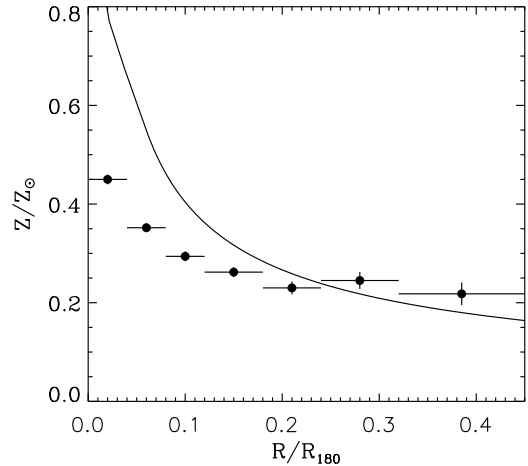}} \\
  \caption{Comparison between our observed mean profile (circles) and the
  one derived from hydrodynamic simulations (solid line) by \cite{fabjan08}.
  Abundances are expressed in \cite{anders89} solar values and radii in
  units of $R_{180}$.}
  \label{fig: cfr sim}
\end{figure}
In Fig.~\ref{fig: cfr sim} we compare LM08 and F08 profiles.
We note differences at both small and large radii.
In the central region, the F08 profile is much more peaked; however, this difference likely
results from two factors: namely that \cite{fabjan08} analyzed only relaxed clusters and that
our data are limited by the \emph{XMM-Newton} PSF.
At large radii, while the F08 profile shows a constant decrease, ours is consistent with
being flat beyond $\approx 0.2 \; R_{180}$.
This discrepancy could be due to issues related to observations, simulations, or both.
More precisely, observations could be affected by unknown systematic effects, and simulations
could underestimate the metallicity at large radii by underestimating possible convective
motions in the ICM, which might be responsible of mixing metals.

\subsection{Comparison with previous works} \label{sec: comp}
We compare our results with those obtained by \cite{grandi04} and \cite{baldi07}.
\citeauthor{grandi04} (hereafter DM04) have analyzed a sample of 21 hot
($\mathrm{k}T \gtrsim 3.5$~keV), nearby ($z \lesssim 0.1$) galaxy clusters observed
with \emph{BeppoSAX}.
\citeauthor{baldi07} (hereafter BA07) have analyzed 12 very hot ($\mathrm{k}T \gtrsim 6$~keV),
intermediate redshift ($0.1 \lesssim z \lesssim 0.3$) clusters observed with \emph{Chandra}.

Comparing results obtained from different works is not trivial; indeed, cluster physical
properties, instrumental characteristics, and data analysis procedures may differ.
Moreover, each author uses his own recipe to derive a scale radius.
We have rescaled DM04 and BA07 profiles by using the standard cosmology
(see Sect.~\ref{sec: intro}) and deriving the scale radius, $R_{180}$, as explained in
Sect.~\ref{sec: mean prof}; the aim is to reduce all inhomogeneities as much as possible.
We also converted DM04 abundances from \cite{grevesse98} to \cite{anders89} solar values.

\begin{figure}
  \centering
  \resizebox{80mm}{!}{\includegraphics[angle=0]{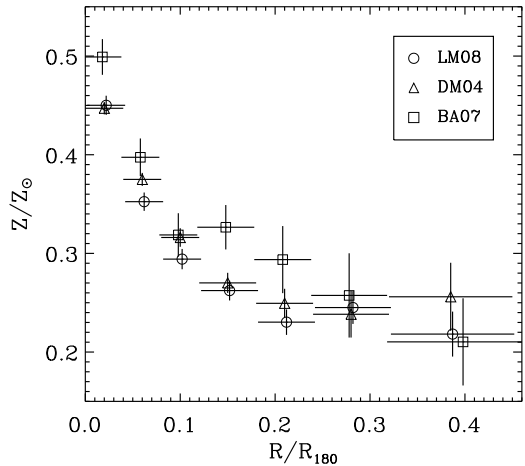}} \\
  \caption{Mean metallicity profiles obtained from this work (LM08,
  circles), by \cite{grandi04} (DM04, triangle), and \cite{baldi07}
  (BA07, squares). Abundances are expressed in \cite{anders89} solar
  values and radii in units of $R_{180}$. The radii have been slightly
  offset in the plot for clarity.}
  \label{fig: Z cfr}
\end{figure}
In Fig.~\ref{fig: Z cfr} we compare LM08, DM04, and BA07 mean metallicity profiles.
Even if cluster samples cover a different redshift range and the instruments (i.e.
\emph{XMM-Newton}, \emph{BeppoSAX}, and \emph{Chandra}) present different characteristics,
the mean profiles are remarkably similar over the entire radial range.

%__________________________________________________________________

\section{Summary and conclusions} \label{sec: concl}
We analyzed a sample of $\approx 50$ hot, intermediate redshift galaxy clusters
to measure their radial properties, in this paper we focused on metallicity profiles.
Our main results are summarized as follows.
\begin{itemize}
   \item The mean metallicity is $0.45 \; Z_\odot$ in the center and decreases out to
	 $\approx 0.2 \; R_{180}$; beyond $0.2 \; R_{180}$ the metallicity is consistent
	 with being flat at $0.23\pm0.01 \; Z_\odot$.
   \item The profiles show a large scatter, which is mostly of statistical origin.
	 In the central regions the scatter (i.e. 22~$\pm$~2\%) is also related to the
	 presence of cool core clusters, in the outer regions it (i.e. 14~$\pm$~8\%)
	 is comparable to systematics.
   \item There is no evidence of profile evolution from $z = 0.1$ to $z = 0.3$.
   \item We find a mean metallicity $\approx 20\%$ lower than found by \cite{maughan08}
	 using \emph{Chandra}, when analyzing roughly the same regions for clusters in the
	 same redshift range.
   \item We point out the existence of a systematic effect, affecting in particular distant
	 clusters, which, if unaccounted for, could cause a false detection of evolution.
   \item When comparing our mean profile to the one derived from hydrodynamic simulations by
	 \cite{fabjan08}, we find differences at small and large radii.
	 In particular, while the profile obtained by \cite{fabjan08} shows a constant decrease,
	 ours is consistent with being flat beyond $\approx 0.2 \; R_{180}$.
   \item When comparing our mean profile to those obtained by recent works with \emph{BeppoSAX}
	 \citep{grandi04} and \emph{Chandra} \citep{baldi07}, we find remarkable agreement
	 over the entire radial range.
\end{itemize}

Our results have been obtained using the same novel data analysis technique as described in
Paper~I; i.e., the background modeling rather than the background subtraction, and the Cash
statistic rather than the $\chi^2$.

%__________________________________________________________________

\begin{acknowledgements}
We acknowledge the financial contribution from contracts ASI-INAF I/023/05/0 and I/088/06/0.
%We thank S.~Ghizzardi, M.~Rossetti, and S.~De~Grandi for a careful reading
%of the manuscript.
We thank A.~Baldi for kindly providing his data.
\end{acknowledgements}

\bibliographystyle{aa}
\bibliography{0113}

\appendix

\section{Discussion of systematics} \label{app}
We made use of montecarlo simulations to test the reliability of our metallicity measurements.
The simulation procedure is similar to that described for the source-only case in Sect.~2 of
\cite{leccardi07}.
We considered a thermal (MEKAL in XSPEC) spectrum only, without a background.
Input parameters are 6~keV temperature, $0.25 \; Z_\odot$ metallicity, 0.2 redshift, and
$9\times10^{-4}$ (in XSPEC units) normalization.
We used the Abell~1689 EPIC-MOS1 observation as a guideline, for producing RMF and ARF, and
for choosing typical input model parameters.
We considered the 3.5$^\prime$-4.5$^\prime$ ring only.
For each channel, we perturbed the number of counts with a Poisson distribution centered on
the expected value, and repeated this procedure 1000 times to obtain 1000 spectra, which
simulate 1000 independent measurements of the source.
We fit simulated spectra with a MEKAL model in the 0.7-10.0~keV energy band using the Cash
statistic.
Temperature, metallicity, redshift, and normalization were allowed to vary within their XSPEC
standard ranges.
Here we focused on the metallicity, for which the standard allowed range is between 0 and
$10^{3}$ in solar units.
For each measurement, we determined the best fit value and the one-sigma uncertainties.

\begin{figure}
  \centering
  \resizebox{80mm}{!}{\includegraphics[angle=0]{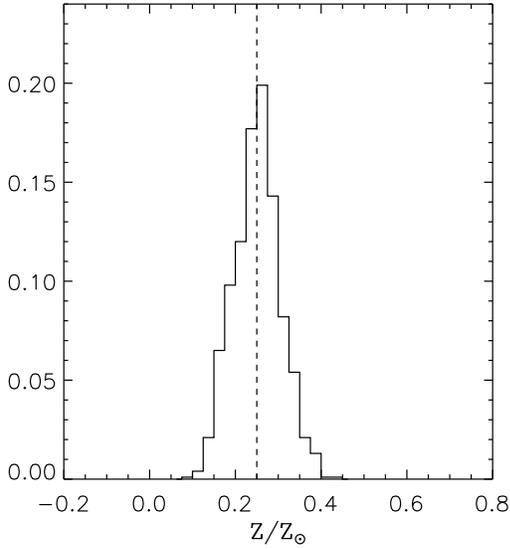}} \\
  \caption{Frequency distribution for metallicity (expressed in solar units)
  best fit values for a 100~ks exposure time. The dashed line indicates the
  input value, i.e. $0.25 \; Z_\odot$.}
  \label{fig: histo100}
\end{figure}
In Fig.~\ref{fig: histo100} we report the frequency distribution for
metallicity best fit values for a 100~ks exposure time (i.e. for spectra
with high statistical quality); as expected, the curve is very similar to a
narrow Gaussian peaked around the input value (i.e. $0.25 \; Z_\odot$).
The mean, the median, and the weighted (over one-sigma uncertainties) average
are all close to $0.25 \; Z_\odot$, namely the weighted average is
$0.246\pm0.002 \; Z_\odot$.
\begin{figure}
  \centering
  \begin{tabular}{cc}
    \hspace{-7mm}
    \resizebox{48mm}{!}{\includegraphics{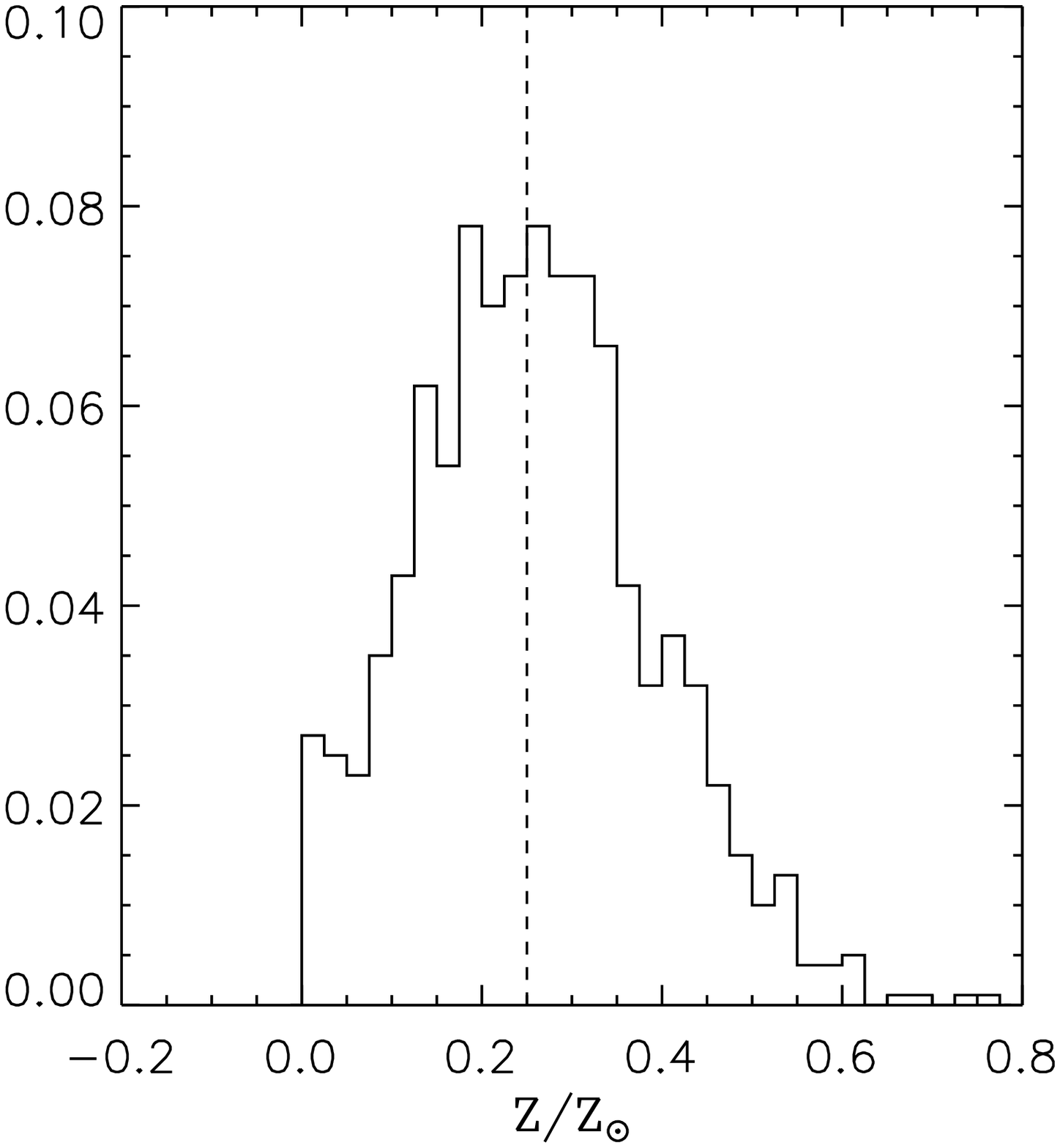}} &
    \hspace{-7mm}
    \resizebox{48mm}{!}{\includegraphics{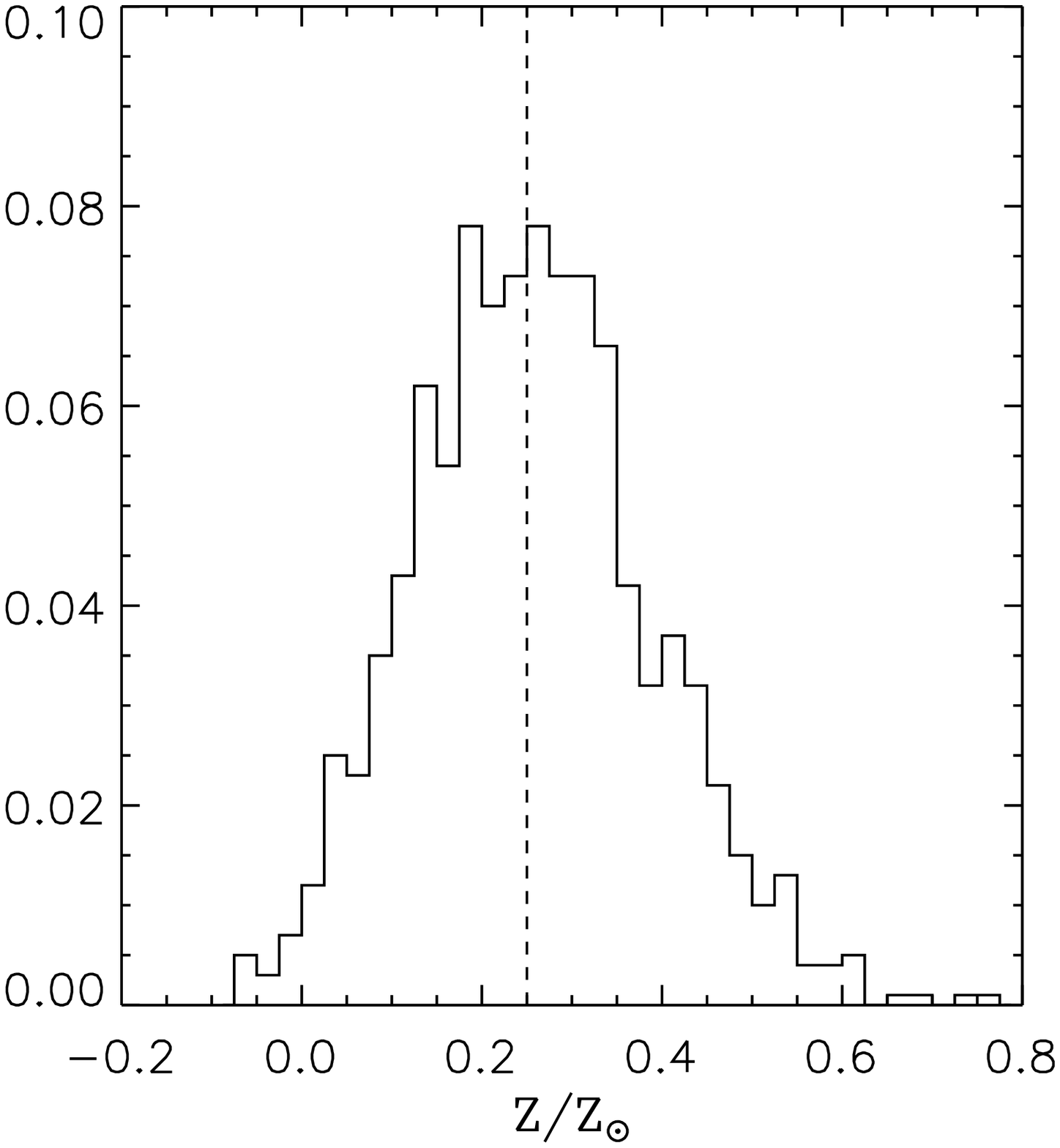}} \\
  \end{tabular}
 \caption{Frequency distribution for metallicity (expressed in solar units)
  best fit values for a 20~ks exposure time. The dashed lines indicate the
  input value, i.e. $0.25 \; Z_\odot$. In the left panel, the metallicity
  is forced to be positive; in the right one, it can assume negative values
  too. The two histograms only differ around zero.}
  \label{fig: histo20}
\end{figure}
In the left panel of Fig.~\ref{fig: histo20} we report the same histogram for
a 20~ks exposure time (i.e. for spectra with standard statistical quality).
The curve is peaked on the input value, but the variance, which mainly depends
on the number of counts around the energy of the emission line, is higher.
Since the metallicity is forced to be positive, the curve is truncated and a
dozen points pile up near the boundary (i.e. zero).
In this case, the mean and the median are close to $0.25 \; Z_\odot$, instead
the weighted average is $0.193\pm0.003 \; Z_\odot$, i.e. $\approx$~25\% lower
than the input value.
Part of this discrepancy is due to a boundary effect.
For measurements characterized by a negative fluctuation of the metallicity,
the minimum of the $\chi^2$ curve lies outside the allowed range.
For these cases, the best fit values are forced to be above zero and the
uncertainties have unreliable small values; therefore, when computing a weighted
average, these measurements have the highest statistical weight, and the net
effect is an underestimate of the real metallicity.
When fitting exactly the same spectra, allowing the metallicity to be negative,
we obtain the histogram reported in the right panel of Fig.~\ref{fig: histo20}.
For this case, the tails of the curve are more symmetric, all measurements have
roughly the same uncertainty, as expected, and the weighted average is
$0.229\pm0.004 \; Z_\odot$, i.e. $\approx$~8\% lower than the input value.
This simple solution allowed us to correct for most of the underestimate; however,
a small (i.e. $\approx 0.02 \; Z_\odot$) systematic still affects our measurements,
especially in the outer regions.
We also performed more realistic simulations, by introducing a background, and
obtained substantially similar results.

Although allowing observables to assume unphysical values is against common sense, there
are measurement procedures that can yield unphysical values.
Deciding to accept only physical values and reject others will clearly result in a bias.
This is a general issue, which does not pertain to astrophysics alone.
An interesting example we found concerns the analysis of data from the Collider Detector
at Fermilab\footnote{http://www-cdf.fnal.gov/physics/statistics/statistics\_faq.html\#ssel1}.
A statistical committee specifically appointed to provide guidelines for the analysis of
the Collider Detector data recommends a treatment similar to the one we propose here, i.e.
unphysical values can be used in statistical procedures.

The boundary effect we have just pointed out should play an important role when
comparing subsamples characterized by a different statistical quality of the data.
For example, in Sect.~\ref{sec: evolution} we compare the metallicity obtained
for near and distant clusters.
In our sample, near cluster spectra usually have a better statistical quality
for various reasons (e.g. longer observations, cosmological dimming effect).
If the metallicity is allowed to vary between 0 and $5 \; Z_\odot$, the mean
metallicity beyond $0.20 \; R_{180}$ is $0.249\pm0.011 \; Z_\odot$ and
$0.188\pm0.014 \; Z_\odot$, for near and distant clusters respectively.
The measured discrepancy of $0.061\pm0.018 \; Z_\odot$ has a significance of more
than three-sigma.
Conversely, if the metallicity is allowed to vary between $\pm5 \; Z_\odot$
(see Sect.~\ref{sec: evolution}), the discrepancy of $0.029\pm0.019 \; Z_\odot$ is
consistent with a purely statistical fluctuation.

We then warn X-ray astronomers about the existence of this kind of systematics,
which could affect the measurement of the metallicity, especially for the case
of low metallicity, statistically poor spectra.

\end{document}